\documentclass[twocolumn,english]{revtex4-1}
\usepackage[T1]{fontenc}
\usepackage[latin9]{inputenc}
\usepackage{amsbsy}
\usepackage{graphicx}

\makeatletter
\usepackage{hyperref}

\makeatother

\usepackage{babel}
\begin{document}
\title{Reveal flocking phase transition of self-propelled active particles
by machine learning regression uncertainty}
\author{Wei-Chen Guo}
\affiliation{Institute for Theoretical Physics, School of Physics, South China
Normal University, Guangzhou 510006, China}
\affiliation{Guangdong Provincial Key Laboratory of Quantum Engineering and Quantum
Materials, Guangdong-Hong Kong Joint Laboratory of Quantum Matter,
South China Normal University, Guangzhou 510006, China}
\author{Bao-Quan Ai}
\email{aibq@scnu.edu.cn}

\affiliation{Institute for Theoretical Physics, School of Physics, South China
Normal University, Guangzhou 510006, China}
\affiliation{Guangdong Provincial Key Laboratory of Quantum Engineering and Quantum
Materials, Guangdong-Hong Kong Joint Laboratory of Quantum Matter,
South China Normal University, Guangzhou 510006, China}
\author{Liang He}
\email{liang.he@scnu.edu.cn}

\affiliation{Institute for Theoretical Physics, School of Physics, South China
Normal University, Guangzhou 510006, China}
\affiliation{Guangdong Provincial Key Laboratory of Quantum Engineering and Quantum
Materials, Guangdong-Hong Kong Joint Laboratory of Quantum Matter,
South China Normal University, Guangzhou 510006, China}
\begin{abstract}
We develop the neural network based \textquotedblleft learning from
regression uncertainty\textquotedblright{} approach for automated
detection of phases of matter in nonequilibrium active systems. Taking
the flocking phase transition of self-propelled active particles described
by the Vicsek model for example, we find that after training a neural
network for solving the inverse statistical problem, i.e., for performing
the regression task of reconstructing the noise level from given samples
of such a nonequilibrium many-body complex system\textquoteright s
steady state configurations, the uncertainty of regression results
obtained by the well-trained network can actually be utilized to reveal
possible phase transitions in the system under study. The noise level
dependence of regression uncertainty assumes a non-trivial M-shape,
and its valley appears at the critical point of the flocking phase
transition. By directly comparing this regression-based approach with
the widely-used classification-based \textquotedblleft learning by
confusion\textquotedblright{} and \textquotedblleft learning with
blanking\textquotedblright{} approaches, we show that our approach
has practical effectiveness, efficiency, good generality for various
physical systems across interdisciplinary fields, and a greater possibility
of being interpretable via conventional notions of physics. These
approaches can complement each other to serve as a promising generic
toolbox for investigating rich critical phenomena and providing data-driven
evidence on the existence of various phase transitions, especially
for those complex scenarios associated with first-order phase transitions
or nonequilibrium active systems where traditional research methods
in physics could face difficulties.
\end{abstract}
\maketitle

\section{Introduction}

In recent years, the machine learning techniques based on the artificial
neural network (ANN) have been increasingly utilized to assist research
in the extensive fields of condensed matter physics and statistical
physics, particularly since the establishing of two pioneering approaches
dubbed \textquotedblleft learning with blanking\textquotedblright{}
\citep{Melko_Nat_Phys_2017,van_Nieuwenburg_Nat_Phys_2017} and \textquotedblleft learning
by confusion\textquotedblright{} \citep{van_Nieuwenburg_Nat_Phys_2017}.
By utilizing the powerful ability of ANNs in classification to identify
phases of matter \citep{Melko_Nat_Phys_2017,van_Nieuwenburg_Nat_Phys_2017,Guo_EPL_2021,Venderley_PRL_2018,Melko_PRB_2018,Lee_PRE_2019,Khatami_PRX_2017,Broecker_Sci_Rep_2017},
these two approaches and their variants have successfully provided
data-driven new evidence on the existence of various phases of matter
in different many-body complex systems, and data-driven estimations
for the critical points of the associated phase transitions. Their
practical successes can now be found all over the fields, including
some tricky scenarios related to non-equilibrium \citep{Guo_EPL_2021,Venderley_PRL_2018},
topological defects \citep{Melko_PRB_2018,Lee_PRE_2019}, and strongly
correlated fermions \citep{Khatami_PRX_2017,Broecker_Sci_Rep_2017},
etc. For both classical and quantum systems, due to the versatility
of ANNs in pattern recognition and data fitting, these machine learning
approaches can readily deal with not only the data generated from
numerical simulations, but also the experimental data \citep{Carrasquilla_AdvPhysX_2020,Deng_FR_2023,Rem_Nat_Phys_2019}.
However, on the other hand, due to the insufficient clarity regarding
the underlying working mechanism of ANNs \citep{Gokmen_PRL_2021,Gokmen_PRE_2021,Kim_Nat_Commun_2021},
they often lack a direct physical connection between the recognized
classes among the raw data analyzed by ANNs and the distinct phases
of matter in the system under study \citep{Gokmen_PRL_2021,Gokmen_PRE_2021,Kim_Nat_Commun_2021},
leaving a challenge to the deeper applications of machine learning
techniques and ANNs in physics.

In the face of this challenge, it is worth noting that the powerful
ability of ANNs is not limited to handling classification tasks. In
fact, ANNs also excel in regression, and significantly, when ANNs
are trained to perform regression tasks, the readily interpretable
meaning of their outputs can usually be traced back to the regression
tasks themselves straightforwardly. For instance, corresponding to
the forward thinking of studying a physical system, i.e., finding
the system's possible states based on system parameter's given values,
the so-called inverse statistical problem (ISP) \citep{Nguyen_Adv_Phys_2017}
refers to the regression task of finding the system parameter's possible
values based on the system's given state. When the ANNs are trained
to perform such an ISP regression task, their outputs are known directly
to be the reconstructed system parameter itself, which is in sharp
contrast to those ambiguous classes encountered in performing classification
tasks. Actually, the connection between the regression results of
ANNs to the conventional notions of physics have already started to
garner attention from physicists. As a case in point, the opportunities
from the symbolic regression \citep{Tegmark_PRE_2021,Tegmark_PRL_2022,Tegmark_PRL_2021,Tegmark_Sci_Adv_2020}
towards automated theory building are being explored, where ANNs are
found to be capable of extracting the equations of motion \citep{Tegmark_PRE_2021},
symmetries \citep{Tegmark_PRL_2022}, conservation laws \citep{Tegmark_PRL_2021}
from various types of data of physical systems, and even 100 equations
in Feynman Lectures on Physics \citep{Tegmark_Sci_Adv_2020}. These
investigations demonstrate that compared to the classification results
of ANNs, the regression results of ANNs have the larger opportunities
to be interpreted in physics.

Hence, there arise the regression-based machine learning approaches
that utilize ANNs for automated detection of phases of matter, such
as the recently-established ``learning from regression uncertainty''
(LFRU) approach \citep{Guo_NJP_2023}. The generic application of
regression uncertainty in learning continuous phase transitions have
been demonstrated on the Ising and $q$-state clock models, and the
intrinsic connection between regression uncertainty and the system's
response properties has been revealed \citep{Guo_NJP_2023}. But further
validation is still required to assess the generality of this new
approach, particularly in the complex scenarios associated with the
first-order phase transition in the non-equilibrium (NEQ) and non-lattice
systems. Is LFRU still effective and efficient? One clearly knows
that in sharp contrast to the continuous phase transition, there is
no diverging correlation length scale associated with the first-order
phase transition. This protects their rich physics at different length
scales from being washed out by the diverging correlation length scale,
but also makes their physics difficult to be studied by powerful tools
such as the renormalization group \citep{Binder_RPR_1987}. It is
known as well that in the NEQ many-body systems, the general absence
of the detailed balance naturally gives rise to much richer physics
comparing to their equilibrium counterparts, e.g., the irregular behavior
of turbulent flows \citep{Falkovich_RMP_2001}, but it also leads
to the situation that no unified approach or even any general guiding
rule seems to exist when one attempts to tackle various physical problems
in the NEQ scenario \citep{Jarzynski_Nat_Phys_2015}. Considering
the versatility of ANNs in pattern recognition and data fitting, they
are well-suited for scenarios like this. Concerning the complex scenarios
associated with the first-order phase transition in active NEQ systems,
if the ANN-based machine learning approaches, e.g., LFRU (and also
the ``learning with blanking'' and ``learning by confusion'' approaches),
can effectively and efficiently handle without the need for additional
case-by-case designs, then they can serve as a promising generic tool
and help to unveil the rich physics in such systems.

In this work, we investigate the generic application of LFRU in learning
first-order phase transitions in active NEQ systems. We address the
question for revealing the flocking phase transition of self-propelled
active particles described by the Vicsek model \citep{Vicsek_PRL_1995,Toner_Ann_Phys_2005,Chate_PRL_2004,Chate_PRE_2008}
for example. This is a stochastic dynamical model with extrinsic noise
(also known as vectorial noise), initially developed to simulate the
flocking behavior of flying birds in low-visibility conditions like
foggy weather, and it also has been one of the foundational models
in the field of statistical physics for self-propelled active particle
systems, exhibiting rich collective dynamics and self-organization
phenomena \citep{Vicsek_PRL_1995,Toner_Ann_Phys_2005,Chate_PRL_2004,Chate_PRE_2008}.
The change in the noise level in this active NEQ system {[}see Fig.~\ref{fig:Samples}
and Fig.~\ref{fig:ISP}(a){]} can drive a first-order phase transition
between the flocking phase (corresponding to low noise level, where
the directions of motion of all particles are generally aligned) and
the disordered phase (corresponding to high noise level, where the
system's rotational symmetry is preserved) \citep{Vicsek_PRL_1995,Toner_Ann_Phys_2005,Chate_PRL_2004,Chate_PRE_2008}.
Here in this work, we find that even in such a complex scenario associated
with a first-order phase transition in an active NEQ system, ANNs
can still be readily trained to perform the ISP regression and successfully
reconstruct the noise level {[}see Fig.~\ref{fig:ISP}(b){]}. Then
we investigate the uncertainty of the regression results obtained
by the ANNs, and find that its noise level dependence contains hidden
information, with the curve assuming an M-shape {[}see Fig.~\ref{fig:Approaches}(a){]}.
Most significantly, we find that this M-shaped curve of regression
uncertainty, which is figured out autonomously by the well-trained
ANNs, can be utilized to reveal the first-order flocking phase transition
of self-propelled active particles in this active NEQ system. The
existence of the non-trivial minimum of regression uncertainty manifests
the presence of the flocking phase transition, and the position of
the non-trivial minimum actually corresponds to the critical noise
level of this flocking phase transition. Building upon our recent
investigations \citep{Guo_NJP_2023}, our findings in this work clearly
demonstrate the good generality of LFRU for various physical systems
across interdisciplinary fields. We also study the practical effectiveness
of the widely-used classification-based \textquotedblleft learning
by confusion\textquotedblright{} and \textquotedblleft learning with
blanking\textquotedblright{} approaches on revealing the same flocking
phase transition of self-propelled active particles, directly compare
LFRU with them on the efficiency, the requirement on prior physical
knowledge, and the possibility of being interpretable via conventional
notions of physics, and discuss these three approaches' similarities
and differences and their respective characteristics.

\section{Flocking phase transition and the inverse statistical problem of
self-propelled active particles\label{sec:ISP}}

\subsection{System and model}

\begin{figure}
\noindent \begin{centering}
\includegraphics[width=3.3in]{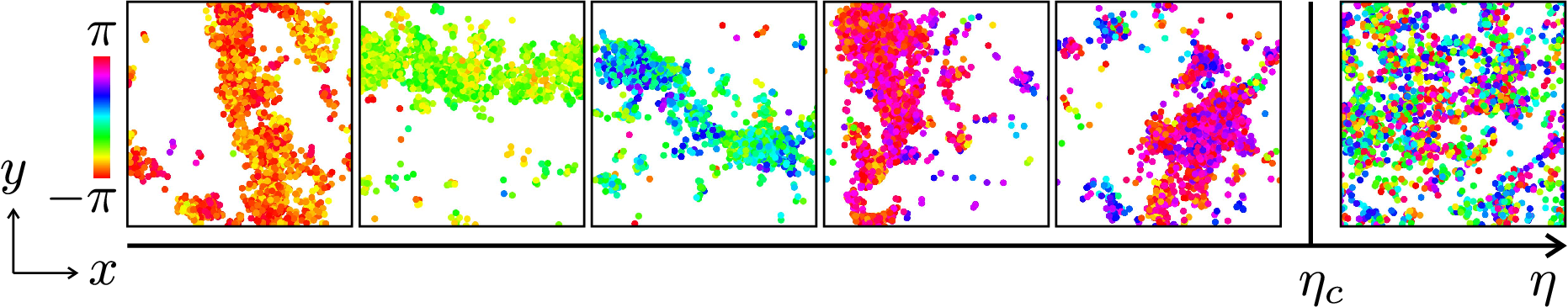}
\par\end{centering}
\caption{\label{fig:Samples}Typical samples corresponding to different noise
levels that are generated by numerical simulations. In every sample,
each of the circular markers represents a single self-propelled particle
in the two-dimensional space, with their spatial distribution representing
the instantaneous spatial distribution of self-propelled particles,
and their color distribution representing the instantaneous angular
distribution of directions of motion of these self-propelled particles.
Among the samples shown here for instance, the five samples in the
left are in the flocking phase, and the rightmost one is in the disordered
phase. See text for more details.}
\end{figure}

The physical system under consideration consists of $N$ self-propelled
active particles in a two-dimensional box of size $L\times L$ with
periodic boundary conditions under influences from environmental fluctuations,
e.g., flying birds in low-visibility conditions like foggy weather.
These self-propelled active particles share a free speed $v_{0}$,
and at any time $t$, each particle $i$ decide to adjust its own
velocity $\boldsymbol{v}_{i}(t)$ according to the average velocity
of all particles located within its neighborhood $U_{i}$ (including
$i$ itself). This can be described by the Vicsek model with extrinsic
noise, where the collective behavior of self-propelled active particles
is modeled by a set of stochastic discrete-time dynamical equations
\citep{Vicsek_PRL_1995,Toner_Ann_Phys_2005,Chate_PRL_2004,Chate_PRE_2008}:

\begin{equation}
\boldsymbol{v}_{i}(t+\Delta t)=v_{0}\vartheta\left(\eta\mathcal{N}_{i}\boldsymbol{\xi}_{i}+\sum_{j\in U_{i}}\boldsymbol{v}_{j}(t)\right).\label{eq:dynamical_equations}
\end{equation}
Here, $\Delta t$ is the discrete time step, $\vartheta$ is a normalization
operator {[}$\vartheta(\boldsymbol{w})=\boldsymbol{w}\slash\vert\boldsymbol{w}\vert${]},
and $\mathcal{N}_{i}$ denotes the number of particles within $U_{i}$.
The neighborhood $U_{i}$ is a disk with radius $r_{0}$ centered
at the location of $i$, and its determination is also subject to
periodic boundary conditions. The random unit vector $\boldsymbol{\xi}_{i}$
is a vectorial noise, with $\eta\in[0,1]$ being the noise level that
captures the environmental fluctuations. A key feature of this system
is that, for the fixed density $\rho\equiv N\slash L^{2}$ of self-propelled
active particles, the change in the noise level $\eta$ can drive
a first-order phase transition between the flocking phase and the
disordered phase, as illustrated in Fig.~\ref{fig:Samples}, which
is characterized by the jump in the system's global group velocity
$\bar{v}=\vert\sum_{j=1}^{N}\boldsymbol{v}_{j}\slash(Nv_{0})\vert$
\citep{Vicsek_PRL_1995,Toner_Ann_Phys_2005,Chate_PRL_2004,Chate_PRE_2008}.
In the following, we focus on the case with $N=2048,\rho=2,L=32,v_{0}=0.5,r=1$
for instance, and in this case, the jump in $\bar{v}$ appears at
$\eta_{c}=0.626\pm0.006$ as one can see from Fig.~\ref{fig:ISP}(a).
The concrete goal of our work is to apply LFRU to make the ANNs automatically
extract this critical noise level $\eta_{c}$ by analyzing the data
of the system's spatial distributions and velocity distributions as
shown in Fig.~\ref{fig:Samples}.

\subsection{Machine learning}

To investigate this active NEQ system of self-propelled active particles,
whether via applying classification-based or regression-based approaches,
one shall first prepare the data in a form that is suitable for analysis
by the ANNs. There are various ways to achieve this, and here for
convenience, we adopt a similar way to the usage of the ANNs in the
image processing applications like facial recognition, i.e., preparing
the data in the form of images. As illustrated in Fig.~\ref{fig:Samples},
in every sample, each of the circular markers represents a single
self-propelled particle in the two-dimensional space, with their spatial
distribution representing the instantaneous spatial distribution of
self-propelled particles, and their color distribution representing
the instantaneous angular distribution of directions of motion of
these self-propelled particles. Since we shall directly employ an
industrially mature deep ANN architecture, known as the residual neural
network (ResNet) \citep{Kaiming_He_Proceedings_IEEE_CVPR_2016}, whose
standard input size is $3\times224\times224$, and thus we accordingly
prepare the samples into images of $224\times224$ pixels (3 for RGB
channels of the images). These samples are then divided into three
categories forming the training dataset, the validation dataset, and
the test dataset.

The so-called training of ANNs refers to traversing the samples in
the training dataset for several epochs, and in each epoch, the ANN
works as a $3\times224\times224\rightarrow1$ map (for the ISP regression
tasks) or $3\times224\times224\rightarrow2$ map (for the binary classification
tasks). The ANN yields one single value (ISP regression) or two values
(binary classification) as outputs concerning every input sample,
and then a loss function is calculated based on these outputs and
the samples' attached labels. According to the back-propagation, the
values of the ANN's numerous trainable parameters (e.g., weights and
biases among neurons) are optimized towards minimizing the loss function.
For the ISP regression task involved in Sec.~III~A in the implementation
of LFRU, the loss function can be the mean squared error between the
reconstructed noise level $\eta_{R}$ and the actual noise level $\eta$
at which the input sample is generated. While for the binary classification
tasks involved in the implementations of \textquotedblleft learning
by confusion\textquotedblright{} and \textquotedblleft learning with
blanking\textquotedblright , then the loss function can be the cross-entropy
function between the confidence outputs and the class labels (we shall
discuss in details in Sec.~III~B and Sec.~III~C about these binary
classification tasks). To improve the generalization ability of the
trained ANN, the values of the ANN's trainable parameters used after
training are not the ones that minimize the loss function on the training
dataset, but those that minimize the loss function on the validation
dataset. Eventually, the well-trained ANN with these final set of
trainable parameters is applied to the test dataset to evaluate its
learning performance. In the following, let us start with discussing
the ANN-based ISP regression.

\subsection{Inverse statistical problem}

The LFRU approach for automated detection of phases of matter utilizes
the regression uncertainty in the ISP. To apply LFRU on investigating
the flocking phase transition of self-propelled active particles,
we shall construct a concrete ISP regression task so that the ANN
can learn to deal with it and thereby expose its regression uncertainty.
Concerning the active NEQ system under consideration, corresponding
to the forward thinking, i.e., finding the system's possible steady-states
(spatial distributions and velocity distributions as shown in Fig.~\ref{fig:Samples})
based on the given noise level $\eta$, one quite natural ISP is finding
the possible value of $\eta$ for every given steady-state of the
system. Since the data are generated by directly simulating the stochastic
discrete-time dynamical equations (\ref{eq:dynamical_equations}),
inevitably there might exist a few samples that are highly similar
but actually generated at different $\eta$. As a result, concerning
all the samples generated at the same noise level $\eta$, their reconstructed
noise level $\eta_{R}$ obtained by ANNs (and any other method) will
not be exactly the same, which thus gives rise to the intrinsic regression
uncertainty $U(\eta)$ for the reconstructed noise levels. Straightforwardly,
we can use the standard deviation of well-trained ANN's outputs to
characterize this regression uncertainty at each $\eta$, whose explicit
form reads:
\begin{equation}
U(\eta)\equiv\sqrt{\langle(\eta_{R}-\langle\eta_{R}\rangle)^{2}\rangle},\label{eq:Uncertainty}
\end{equation}
where $\langle\cdot\rangle$ denotes the average over all the samples
generated at $\eta$ in the test dataset. Reducing this intrinsic
regression uncertainty is usually regarded as one of the central goals
of the ISP itself, but here it is just a necessary intermediate product
for revealing possible phase transitions, and theoretically it cannot
be reduced to zero, so we are not intending to pursue its minimization.
Further noticing that this is a complex scenario associated with a
first-order phase transition in an active NEQ system, how to effectively
and efficiently realize the ISP in the Vicsek model is already a non-trivial
open question. The traditional research methods majorly focus on the
inverse Ising problem, yet they usually resort to some case-by-case
methods such as mean-field \citep{Nguyen_PRL_2012} and maximum likelihood
estimation \citep{Periwal_PRE_2020}. In this work, we directly utilizes
the ANN-based machine learning techniques to perform the ISP regression.

\begin{figure}
\noindent \begin{centering}
\includegraphics[width=3.3in]{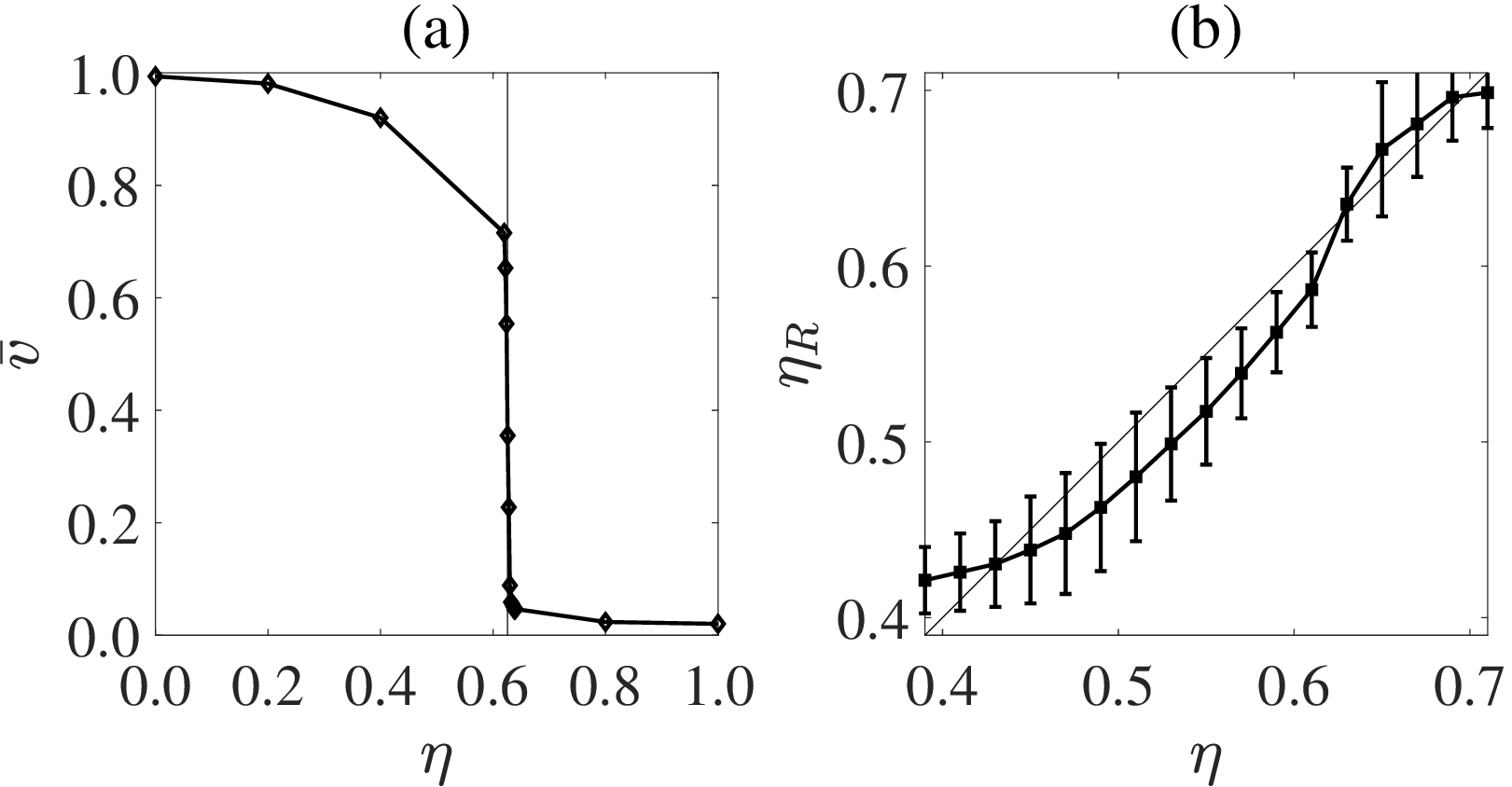}
\par\end{centering}
\caption{\label{fig:ISP}Inverse statistical problem in a self-propelled active
particle system: (a) Noise level dependence of the system\textquoteright s
global group velocity, whose jump at $\eta_{c}=0.626\pm0.006$ characterizes
the first-order flocking phase transition; (b) Noise level dependence
of the reconstructed noise level predicted by the well-trained ANN.
The error bars represent the regression uncertainty $U(\eta)$, and
the diagonal line represent the ideal regression result $\eta_{R}=\eta$.
See text for more details.}
\end{figure}

\begin{figure}
\noindent \begin{centering}
\includegraphics[width=3.3in]{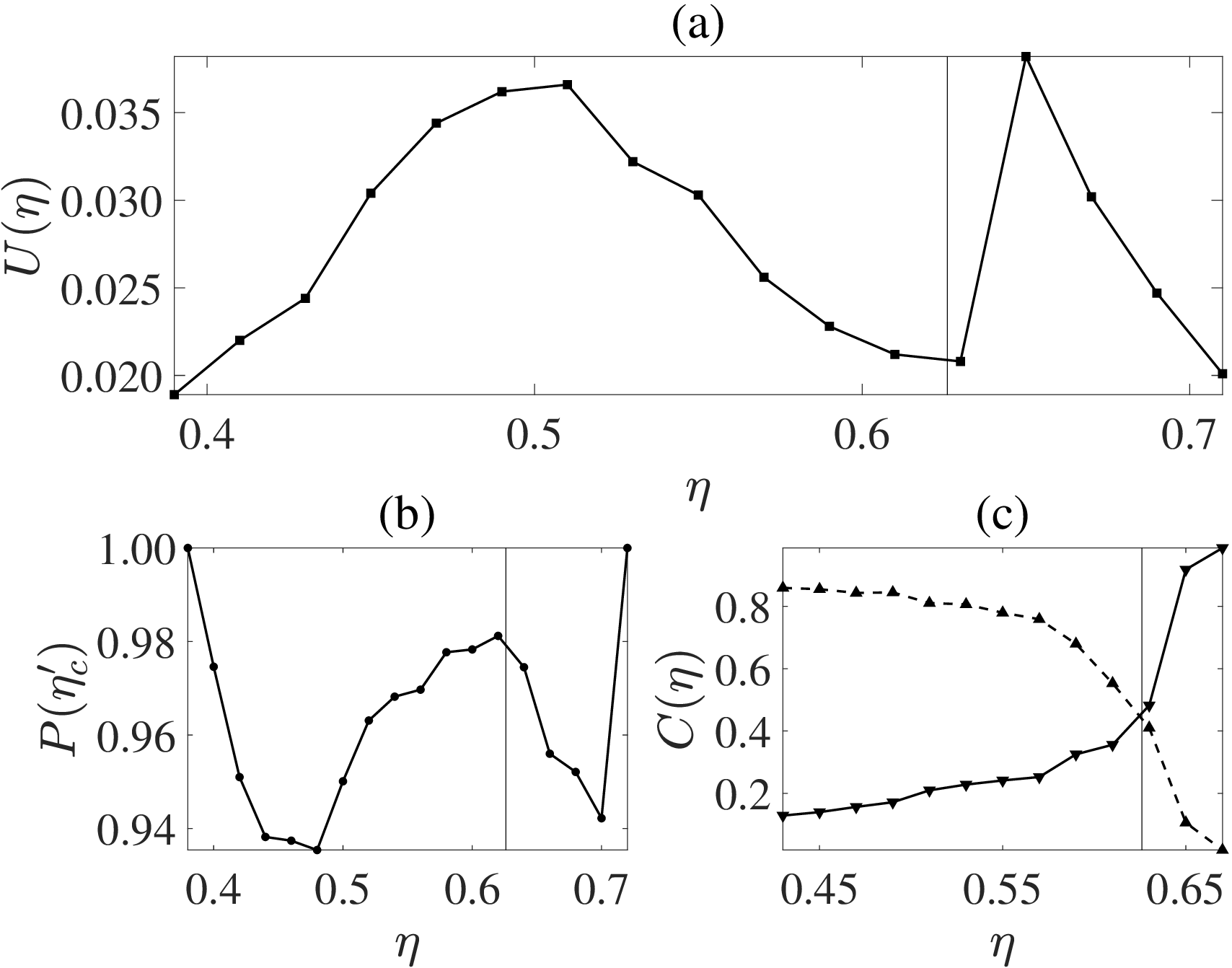}
\par\end{centering}
\caption{\label{fig:Approaches}Revealing the flocking phase transition of
self-propelled active particles via applying three different machine
learning approaches: (a) The LFRU approach; (b) The \textquotedblleft learning
by confusion\textquotedblright{} approach; (c) The \textquotedblleft learning
with blanking\textquotedblright{} approach. See text for more details.}
\end{figure}

\section{Automated extraction of critical noise level\label{sec:Results}}

\subsection{Learning from regression uncertainty (LFRU)}

The datasets involve 17 different noise levels in the range $\eta\in[0.39,0.71]$
with a constant spacing $\Delta\eta=0.02$. For each noise level $\eta$,
there are 2000 samples in the training dataset, 500 samples in the
validation dataset, 2500 samples in the test dataset. The $3.4\times10^{4}$
training samples corresponding to 17 different $\eta$ are together
fed to the ANN, traversing for 20 epochs in the training process.
After the training and the corresponding validation, the learning
performance of the trained ANN is evaluated with the test samples.
Fig.~\ref{fig:ISP}(b) shows the average regression results of 20
well-trained ANNs (all are ResNet, but independently trained and validated).
As one can see from Fig.~\ref{fig:ISP}(b), the reconstructed noise
level $\eta_{R}$ predicted by the well-trained ANN does not exactly
match the actual noise level $\eta$, but almost. This suggests that
ANNs can indeed learn the noise level $\eta$ in this active NEQ system,
overcoming potential interferences such as metastable states, and
hence the outputs of the well-trained ANNs can naturally be considered
as having a direct physical connection to the noise level $\eta$.
Since the automated detection of the flocking phase and the extraction
of the critical noise level $\eta_{c}$ are further built upon the
outputs of the well-trained ANNs, they thus also have the great possibility
of being interpretable via conventional notions of physics.

After verifying that ANNs can deal with the ISP regression in the
active NEQ system under consideration, let us take a deeper look at
the regression uncertainty $U(\eta)$, i.e., the error bars of the
regression results shown in Fig.~\ref{fig:ISP}(b). Its non-trivial
information is actually ``hidden'' in Fig.~\ref{fig:ISP}(b), but
when the noise level dependence of regression uncertainty $U(\eta)$
is plotted explicitly in Fig.~\ref{fig:Approaches}(a), one can clearly
notice that the curve assumes an M-shape, and the valley position
$\tilde{\eta}_{c}=0.63\pm0.01$ is not located at the middle of the
parameter region $[0.39,0.71]$, but automatically corresponds to
the critical noise level $\eta_{c}$ of the system (the vertical lines
in Fig.~\ref{fig:Approaches} represent the critical noise level
$\eta_{c}=0.626\pm0.006$ obtained via traditional methods, i.e.,
by the jump in $\bar{v}$). This suggests that ANNs can successfully
extract the critical noise level $\eta_{c}$ of the flocking phase
transition of self-propelled active particles.

These findings are consistent with our recent investigations that
establish LFRU and demonstrate it on the Ising and $q$-state clock
models \citep{Guo_NJP_2023}, reflecting the LFRU's good generality
for various physical systems across interdisciplinary fields. To summarize,
by utilizing the powerful ability of ANNs in regression and its direct
physical connection to conventional notions of physics, physicists
applying LFRU simply need to provide ANNs with the actual parameter
values of each sample to train for the ISP regression tasks. Then
the uncertainty of regression results obtained by the well-trained
ANNs can be utilized to reveal possible phase transitions in the system
under study. If there is only one distinct phase of matter, i.e.,
no phase transition existing in the parameter region examined by LFRU,
the curve of the regression uncertainty is demonstrated to exhibit
only one trivial peak, without any non-trivial minimum. Once the curve
of the regression uncertainty assumes the M-shape, it reveals that
a phase transition is expected to exist in the examined parameter
region, and the critical point of the phase transition can be extracted
from the valley position of the regression uncertainty.

\subsection{Learning by confusion}

In the following, we shall also apply the widely-used classification-based
\textquotedblleft learning by confusion\textquotedblright{} and \textquotedblleft learning
with blanking\textquotedblright{} approaches on revealing the same
flocking phase transition of self-propelled active particles as a
direct comparison. To train ANNs for the classification tasks (more
specifically, binary classification tasks here in this case), compared
to the usage above, one shall change the loss function into another
candidate that is suitable for classification tasks instead of regression
tasks, e.g., the cross-entropy function. Moreover, here each sample
shall be attached with a two-valued class label $(C_{1},C_{2})$ which
can be interpreted as probabilities, with the label $(1,0)$ indicating
that the sample is $100\%$ likely to be class-A, the label $(0,1)$
indicating that the sample is $100\%$ likely to be class-B. Accordingly,
the two outputs of the ANN for each sample can also be interpreted
as probabilities. For instance, an output $(0.6,0.4)$ means that
the ANN has a $60\%$ confidence to recognize the sample as class-A,
and a $40\%$ confidence to recognize the sample as class-B. Naturally,
the classification result given by the ANN is class-A if $C_{1}>C_{2}$,
and class-B if $C_{1}<C_{2}$ .

Now let us start with the \textquotedblleft learning by confusion\textquotedblright{}
approach. This approach reveals possible phase transitions via monitoring
the contrast between the ANN's good and bad recognition performance
when confusing labels are deliberately attached to some of the samples.
To implement it, one shall purpose an arbitrary noise level $\eta_{c}^{\prime}$,
declaring that any sample satisfying $\eta<\eta_{c}^{\prime}$ ($\eta>\eta_{c}^{\prime}$)
belongs to class-A (class-B). In general, this binary classification
rule is for no reason at all, and these ambiguous classes are entirely
unrelated to the distinct phases of matter in the system under study
(e.g., the flocking phase and the disordered phase). After training,
the ANN is applied to the test dataset to evaluate its learning performance
concerning this binary classification task associated with $\eta_{c}^{\prime}$.
In the $m$ test samples generated at different noise levels, when
the trained ANN match $m^{\prime}$ samples of them with their attached
class labels, its recognition performance is estimated by the classification
accuracy $P(\eta_{c}^{\prime})=m^{\prime}\slash m$. Repeating the
above procedure with a series of different purposed $\eta_{c}^{\prime}$,
one can thus establish the $\eta_{c}^{\prime}$ dependence of the
classification accuracy $P(\eta_{c}^{\prime})$.

As one can see from Fig.~\ref{fig:Approaches}(b), the curve of the
classification accuracy $P(\eta_{c}^{\prime})$ assumes a W-shape.
Noticing that except the two trivial choices $\eta_{c}^{\prime}=0$
and $\eta_{c}^{\prime}=1$ which indicate essentially no classification
is performed, for any arbitrary $\eta_{c}^{\prime}$ that does not
match the physical flocking transition point $\eta_{c}$, its corresponding
way of labeling the samples will inevitably confuse the ANN in the
training process by those wrong labels, and hence lowers the classification
accuracy $P(\eta_{c}^{\prime})$ of the trained ANN.

To see this, let us concretely discuss the $\eta_{c}^{\prime}>\eta_{c}$
case for example (the $\eta_{c}^{\prime}<\eta_{c}$ case is just similar).
Those class-A samples satisfying $\eta_{c}<\eta<\eta_{c}^{\prime}$
and the class-B samples satisfying $\eta>\eta_{c}^{\prime}$ are in
fact both in the disordered phase, but opposite class labels have
been attached with them. How could the ANN learn the actually non-existent
``difference'' between these two classes? Meanwhile, they look indeed
different from those other class-A samples satisfying $\eta<\eta_{c}$.
Then how could the ANN learn this actually non-existent ``similarity''?
These confusing labels that deviate from physical facts inevitably
limit the recognition performance of the ANN, and hence the classification
accuracy $P(\eta_{c}^{\prime})$ would not be quite ideal. Naturally,
for the same datasets, the closer $\eta_{c}^{\prime}$ is to the physical
flocking transition point $\eta_{c}$, the fewer such confusing labels
exist, leading to the relatively higher $P(\eta_{c}^{\prime})$. This
means that the classification accuracy $P(\eta_{c}^{\prime})$ is
expected to assume a non-trivial maximum exactly when $\eta_{c}^{\prime}=\eta_{c}$.
Therefore, the ``learning by confusion'' approach takes the peak
position of the W-shaped curve of the classification accuracy $P(\eta_{c}^{\prime})$
as the predicted critical noise level.

In Fig.~\ref{fig:Approaches}(b), the average results of 20 well-trained
ANNs suggest that the maximum of $P(\eta_{c}^{\prime})$ corresponds
to $\tilde{\eta}_{c}=0.62\pm0.01$, which matches well with the critical
noise level $\eta_{c}=0.626\pm0.006$ obtained via traditional methods
(see also the vertical lines in Fig.~\ref{fig:Approaches}). These
investigations manifest that the ``learning by confusion'' approach
also holds its practical effectiveness for extracting the critical
noise level $\eta_{c}$ of the flocking phase transition of self-propelled
active particles by analyzing the data of this active NEQ system's
spatial distributions and velocity distributions as shown in Fig.~\ref{fig:Samples}.

\subsection{Learning with blanking}

Now let us switch to the ``learning with blanking'' approach. This
approach reveals possible phase transitions via directly utilizing
the ANN's ability to identify various phases of matter. When all the
samples are attached with appropriate labels that are consistent with
the physical facts (i.e., the $\eta_{c}^{\prime}=\eta_{c}$ case above),
even though the ANN is trained with only the samples corresponding
to low and high noise levels, blanking the intermediate noise levels,
the ANN is still expected to easily accomplish its binary classification
task. More specifically, here in the training and the validation datasets,
the samples corresponding to $\eta=0.39,0.41$ ($\eta=0.69,0.71$)
are labeled as class-A (class-B), and the other samples corresponding
to the intermediate noise levels $0.41<\eta<0.69$ are not involved
before the ANN is well-trained. After training, the ANN's recognition
confidences are evaluated concerning the test dataset within $[0.43,0.67]$
for instance.

As one can see from Fig.~\ref{fig:Approaches}(c), which shows the
average confidences of 20 well-trained ANNs (all are also ResNet,
but independently trained and validated) for identifying the samples
in $\eta\in[0.43,0.67]$ into class-A or class-B. The dashed and solid
curves in Fig.~\ref{fig:Approaches}(c) represent the noise level
$\eta$ dependence of the class-A confidence $C_{1}$ and the class-B
confidence $C_{2}$, respectively. The intersection point of these
two curves locates at $\tilde{\eta}_{c}\approx0.625$, which is the
critical noise level predicted by the ANN via this approach, since
at the critical point, the instantaneous states of the system can
be either in the flocking phase or in the disordered phase, resulting
in the equal confidences of the well-trained ANN $C_{1}(\eta)=C_{2}(\eta)$.
This predicted critical noise level also matches the one $\eta_{c}=0.626\pm0.006$
obtained via traditional methods (see also the vertical lines in Fig.~\ref{fig:Approaches}),
manifesting that the ``learning with blanking'' approach holds its
practical effectiveness as well.

\section{Comparison}

So far, we have already witnessed in Fig.~\ref{fig:Approaches} that
without additional case-by-case designs for such a complex scenario
associated with a first-order phase transition in an active NEQ system,
the regression-based LFRU approach and two classification-based approaches
can all be readily applied to utilize the powerful ability of ANNs
for extracting the critical noise level $\eta_{c}$. In the following,
let us further discuss their respective characteristics.

\subsection{Efficiency}

The efficiency is a fundamental demand for any practical research
method. Comparing the ANNs performing the regression and classification
tasks, they have just a slight difference in their network architecture,
lying in the number of output neurons (one neuron for the ISP regression,
and two neurons for the binary classification), which thus leads to
their almost equal computational complexity for traversing the same
dataset once. Their loss functions' contributions to the computational
complexity are also roughly the same, and their convergence speeds
are close as well \citep{Guo_NJP_2023}. Consequently, in the implementations
of LFRU and the ``learning by confusion'' approach, the time required
to give birth to a well-trained ANN is approximately equal. However,
a well-trained ANN for the ``learning by confusion'' approach can
just give a single value of $P(\eta_{c}^{\prime})$, and it must train
a series of ANNs examining a series of different $\eta_{c}^{\prime}$
so as to locate the maximum of $P(\eta_{c}^{\prime})$. While in sharp
contrast, when applying LFRU, a complete curve of $U(\eta)$ is directly
obtained from one well-trained ANN. This makes the total time required
of LFRU for automated detection of phases of matter is less than the
total time required of the ``learning by confusion'' approach. As
for the ``learning with blanking'' approach, it requires the least
time since it only involves a small part of the whole datasets, but
this comes at a cost, as it requires prior physical knowledge to a
much greater extent and thus cannot independently realize the automated
detection of phases of matter.

\subsection{Requirement on prior physical knowledge}

To realize the automated detection of phases of matter, any practical
research method based on machine learning techniques is naturally
not expected to require prior physical knowledge on the phases of
matter in the system under study. This topic is related to the concept
of supervision in machine learning. In the terminology of machine
learning, a supervised learning algorithm refers to a machine learning
algorithm that involves the attached labels for each sample as the
target for learning. In this sense, all the three approaches investigated
in this work are technically the supervised learning algorithms. However,
as we have seen in Sec.~\ref{sec:ISP}~C and Sec.~\ref{sec:Results}~A,
when applying LFRU to reveal the flocking phase transition of self-propelled
active particles, the labels are the actual noise level $\eta$ at
which each sample is generated, while the physical target of applying
LFRU is not to solve the ISP itself, but to extract the critical noise
level of the flocking phase transition. The labels provide the ANN
with knowledge about the ISP, having nothing to do with the phases
of matter and possible phase transitions in this system. Therefore,
concerning the applications of machine learning techniques in physical
researches, LFRU can be regarded as unsupervised. The ``learning
by confusion'' approach is also unsupervised in the same sense \citep{van_Nieuwenburg_Nat_Phys_2017},
since the labels are attached according to an arbitrarily purposed
noise level $\eta_{c}^{\prime}$. But it is also noteworthy that its
binary classification implies the prior judgment that there are at
most two phases existing in the system, making it requires further
modifications before it can be applied to deal with those complex
many-body systems with distinct intermediate phases \citep{Lee_PRE_2019}.
And on the other hand, the ``learning with blanking'' approach takes
the intersection point of two confidence curves as the system's phase
transition point, which assumes that there exists and only exists
one phase transition. This makes it not only hard to readily handle
the systems with distinct intermediate phases, but also face great
challenges from the phase coexistence and crossover scenarios, since
a phase transition, a phase coexistence, and a crossover can all similarly
lead to the intersection \citep{Melko_Nat_Phys_2017}. In general,
among these three approaches, the ``learning with blanking'' requires
prior physical knowledge most, followed by the ``learning by confusion''
approach, and LFRU shows a relative advantage in this aspect.

\subsection{Possibility of being interpretable via conventional notions of physics}

Another fundamental demand for automated detection of phases of matter
is the possibility of being physically interpretable. Due to the insufficient
clarity regarding the underlying working mechanism of ANNs \citep{Gokmen_PRL_2021,Gokmen_PRE_2021,Kim_Nat_Commun_2021},
the interpretability of the ANN-based machine learning techniques
themselves is beyond the scope of our work, and here in the implementations
of all the three approaches, we consider the employed ANNs as black-box
maps. But under such circumstances, one can still look for the physical
interpretability of machine learning results obtained by utilizing
these black-box maps, connecting the results to the conventional notions
of physics. In this context, the final step of extracting the critical
noise level via the classification-based approach is completed by
the human rather than by ANNs, i.e., to directly consider the data
boundary point between class-A and class-B as the physical phase transition
point between the flocking phase and the disordered phase. This is
actually injecting additional physical knowledge about this flocking
phase transition into the ANNs afterwards, which weakens the unsupervised
nature of the approach since for the practical application scenarios,
i.e., for an unexplored system, it needs further reasons for this
association. In distinction, ANNs learn faithfully the noise level
$\eta$ of the system under study, and based on this, the results
of the regression-based LFRU can be naturally connected to conventional
notions of physics \citep{Guo_NJP_2023}. The outputs of the well-trained
ANNs are the reconstructed system parameter $\eta$ itself, and hence
the statistical properties of these outputs (such as the regression
uncertainty) are a reflection of the system's statistical properties.
When the statistical properties of the outputs of ANNs exhibit non-trivial
behaviors, for instance, when the regression uncertainty $U(\eta)$
assumes a non-trivial minimum at a certain noise level $\eta=\tilde{\eta}_{c}$,
there is reason to expect that the system's statistical properties
exhibit non-trivial behaviors as well at that point. This thus provides
the justification for taking the noise level $\tilde{\eta}_{c}$ that
corresponds to the non-trivial minimum of regression uncertainty as
the critical noise level of the flocking phase transition in this
active NEQ system. Moreover, noticing that the intrinsic connection
between regression uncertainty and the system's response properties
has been revealed in the Ising and $q$-state clock models \citep{Guo_NJP_2023},
the analogues of such a connection could also exist in the Vicsek
model. These possibilities of numerically and theoretically connecting
the machine learning results of ANNs to the conventional notions of
physics are not often available for the classification-based approaches.

\section{Conclusions}

In this work, after training the ANNs to perform the ISP regression
in the active NEQ system of self-propelled active particles described
by the Vicsek model, we find that the regression uncertainty of the
well-trained ANNs actually contains hidden information that can be
utilized to reveal the flocking phase transition in this system. The
noise level $\eta$ dependence of regression uncertainty $U(\eta)$
assumes an M-shape, providing data-driven new evidence on the existence
of this phase transition. Its valley position provides a data-driven
estimation $\tilde{\eta}_{c}=0.63\pm0.01$ for the critical noise
level, which matches well with the value $\eta_{c}=0.626\pm0.006$
obtained via traditional methods. We further find that the regression-based
approach LFRU developed in this work for automated detection of phases
of matter in active NEQ systems has several distinctive characteristics
and practical advantages. LFRU can complement with the widely-used
classification-based \textquotedblleft learning by confusion\textquotedblright{}
and \textquotedblleft learning with blanking\textquotedblright{} approaches
to serve as a promising generic toolbox, bringing a new perspective
for investigating rich critical phenomena utilizing the ANN-based
machine learning techniques, for those complex scenarios associated
with the first-order phase transitions in active NEQ systems where
traditional research methods in physics could face difficulties. Owing
to the powerful ability of ANNs in regression and its direct physical
connection to conventional notions of physics, these findings could
inspire and guide the further revealing of the connection in the Vicsek
model and Vicsek-like models between regression uncertainty and the
system's statistical properties such as the response properties. In
these regards, we believe that the development of LFRU will stimulate
further efforts in both developing and applying physically interpretable
machine learning approaches to unveil new physics in these active
NEQ systems.
\begin{acknowledgments}
This work was supported by the National Science Foundation of China
(Grants Nos.~12275089, 12075090), the Guangdong Basic and Applied
Research Foundation (Grant Nos.~2023A1515012800, 2022A1515010449),
and the National Key Research and Development Program of China (Grant
No.~2022YFA1405304).
\end{acknowledgments}

\end{document}